\documentstyle[preprint,aps]{revtex}
\begin{document}
\draft
\preprint{}
\title{Dynamics of An Underdamped Josephson Junction Ladder}
\author{Seungoh Ryu \\ Wenbin Yu \\ D. Stroud}
\address{Department of Physics, Ohio State University, Columbus, OH 43210}
\date{\today}
\maketitle
\begin{abstract}
We show analytically that the dynamical equations for an underdamped ladder of
coupled small Josephson junctions can be approximately reduced to the
discrete sine-Gordon equation.  As numerical confirmation,
we solve the coupled Josephson equations for such a ladder in a
magnetic field.  We obtain discrete-sine-Gordon-like IV characteristics,
including a flux flow and  a ``whirling'' regime at low and high currents,
and voltage steps which represent a lock-in between the vortex motion and
linear ``phasons'', and which are quantitatively predicted by a simple
formula.  At sufficiently high anisotropy,
the fluxons on the steps propagate ballistically.
\end{abstract}
\pacs{PACS numbers:74.40.+k, 74.50.+r, 03.20.+i, 05.45.+b, 85.25.C}

\narrowtext

The discrete sine-Gordon equation has been used by several groups to model
so-called hybrid Josephson ladder arrays\cite{ustinov93,strogatz95,zant95}.
Such an array consists of a ladder
of parallel Josephson junctions which are inductively coupled
together, e.\ g.\ by superconducting wires\cite{nakajima}.
The sine-Gordon equation then describes the phase differences
across the junctions.  In an applied magnetic field,
this equation predicts remarkably complex behavior,
including flux flow resistance below a certain critical current,
and a field-independent resistance above that current arising from
so-called ``whirling" modes\cite{strogatz95}.  In the flux flow regime,
the fluxons in this ladder propagate as localized solitons, and
the IV characteristics exhibit
voltage plateaus arising from the locking of solitons
to linear ``spin wave'' modes.   At sufficiently large
values of the  anisotropy parameter $\eta_J$ defined later,
the solitons may propagate ``ballistically'' on the plateaus,
i.\ e.\ may travel a considerable distance even after the
driving current is turned off.

In this Letter, we show that this behavior is all found in a model
in which the ladder is treated as a network
of coupled small junctions arranged along both the edges and the rungs of
the ladder.  This model is often used to treat
two-dimensional Josephson networks\cite{shenoy}, and includes
{\em no} inductive coupling
between junctions, other than that produced by the other junctions.
To confirm our numerical results, we derive
a discrete sine-Gordon equation from our coupled-network model.
Thus, these seemingly quite different
models produce nearly identical behavior for ladders.
By extension, they suggest that some properties of
two-dimensional arrays might conceivably
be treated by a similar simplification.  In
simulations\cite{bobbert,geigenmuller93,yu94,shea},
underdamped arrays of this type show some similarities
to ladder arrays, exhibiting the analogs of both
voltage steps and whirling modes.

Our ladder consists of coupled
superconducting grains, the i$^{th}$ of which has
order parameter $\Phi_i = \Phi_0e^{i\theta_i}$.
Grains $i$ and $j$ are
coupled by resistively-shunted Josephson junctions (RSJ's) with
current $I_{ij}$, shunt resistance $R_{ij}$ and shunt capacitance
$C_{ij}$, with periodic boundary conditions (see Fig. 1).

The phases $\theta_i$ evolve
according to the coupled RSJ equations
$\hbar\dot{\theta_i}/(2e) = V_i$,
$\sum_j M_{ij} \dot{V_j} = I^{ext}_i/I_c-
\sum_j(R/R_{ij})(V_i - V_j)
- \sum_j (I_{ij}/I_c)\sin(\theta_{ij} - A_{ij})$.
Here the time unit is $t_0 = \hbar/(2eRI_c),$
where $R$ and $I_c$ are the shunt resistance and critical current
across a junction in the x-direction (see Fig.\ 1); I$^{ext}_i$ is the external
current fed into the i$^{th}$ node;
the spatial distances are given
in units of the lattice spacing $a$, and
the voltage $V_i$ in units of $I_cR$.
$M_{ij} = - 4\pi e C I_c R^2 /h$ for
$i \neq j,$ and
$M_{ii} = -\sum_{j\neq i}M_{ij}$,
where $C$ is the intergrain capacitance.
Finally, $A_{ij} = (2\pi/\Phi_0)\int_i^j{\bf A}\cdot{\bf dl}$, where
${\bf A}$ is the vector potential.  We assume N plaquettes in the array,
and postulate a current $I$ uniformly injected into each node on the outer
edge and extracted from each node on the inner edge of the ring.  We also
assume a uniform transverse magnetic field
$B \equiv f\phi_0/a^2$,
and use the Landau gauge ${\bf A} = -Bx\, \hat{{\bf y}}$.
We solve these equations
numerically using a fourth-order Runge-Kutta algorithm.

We now show that this model reduces approximately to a
discrete sine-Gordon equation for the {\em phase differences}.
Label each grain
by $(x,y)$ where $x/a = 0, \ldots, N-1$ and $y/a=0, 1$.
Subtracting the equation of motion for
$\theta(x,1)$ from that for $\theta(x,2)$,
and defining
$\Psi(x) = \frac{1}{2}[\theta(x,1) + \theta(x,2)]$,
$\chi(x) = [\theta(x,2)-\theta(x,1)]$, we obtain
a differential equation for $\chi(x)$ which is second-order in time.
This equation may be further simplified using the facts that
$A_{x,y;x\pm 1,y} = 0$ in the Landau gauge, and that
$A_{x,1;x,2}=-A_{x,2;x,1}$, and by defining the discrete Laplacian
$\chi(x+1)-2\chi(x)+\chi(x-1)=\nabla^2\chi(x)$.
Finally, using the boundary conditions,
$I^{ext}(x,2)$ =  - $I^{ext}(x,1) \equiv I$, and introducing
$\varphi(x) = \chi(x) - A_{x,2;x,1}$,
we obtain
\begin{eqnarray}
\label{eqx7}
&[& 1 - \eta_c^2 \nabla^2 ] \beta \ddot{\varphi} =
i - [ 1 - \eta_r^2 \nabla^2 ] \dot{\varphi} -  \sin(\varphi) +  2 \eta^2_J
\times
\nonumber \\
& & \sum_{i=\pm 1}\cos\{ \Psi(x)\!-\!\Psi(x+i)\}
\sin\{[ \varphi(x)\!-\! \varphi(x+i)]/2\},
\end{eqnarray}
where we have defined a dimensionless current $i=I/I_{cy}$, and
anisotropy factors $2 \eta_r^2 = R_y/R_x$, $2 \eta_c^2 = C_x/C_y,$ and
$2 \eta_J^2 = I_{cx}/I_{cy}.$

We now neglect all combined space and time derivatives of order three
or higher.  Similarly, we set the cosine factor equal to unity(this is also
checked numerically to be valid {\em a posteriori})
and linearize the sine factor in the last term, so that the final summation
can be expressed simply as $\nabla^2\phi$.  With these
approximations, eq. (1) reduces to
{\em discrete driven sine-Gordon equation with dissipation}:
\begin{equation}
\label{eqx9}
\beta \ddot{\varphi}+
\dot{\varphi}+ \sin(\varphi) - \eta^2_J \nabla^2 \varphi = i,
\end{equation}
where $\beta = 4\pi eI_{cy}R_y^2C_y/h$.

To confirm the accuracy of this reduction,
we have solved the coupled Josephson equations
on an $8 \times 1$ (N=8) ring.  Fig.\ 1
shows the resulting $IV$ characteristics with $\beta = 33$,
$\eta_J = 0.71$, and several values of $f$.
$\langle V \rangle$ denotes the space and time-averaged voltage
differences across the y-junctions.
There are two regimes.  The first is the ``flux flow regime'' where
$\langle V \rangle (I, f)$ is roughly proportional to $f$ (up to about
$f$=3/8) and to $I$.
In this regime, for each $f$, $\langle V \rangle (I, f)$ also exhibits
a series of voltage steps (see below).
The second regime corresponds
to ``resistance steps'' in which $\langle V \rangle = RI$ independent of
$f$.  This regime is dominated by the whirling modes, and is also discussed
further below.

Fig.\ 2 shows an expanded low-current regime for $f=1/8$,
$\beta = 61$, and several $\eta_J$'s.  The voltage steps are very prominent.
The $IV$ characteristics are hysteretic on each of the steps, as shown with
broken lines for $\eta_J^2 = 1.25$.
Similar hysteretic steps are well known in numerical studies of the
discrete sine-Gordon equation\cite{ustinov93,strogatz95}.
The critical current for the onset
of voltage varies from about $0.2I_c$ at $\eta_J$ = 0.71 to $\approx 0$
for $\eta_J > 1$, in the less discrete regime, similar to results obtained
in \cite{zant95}.

\underline{Soliton behavior.}  In the absence of damping and driving,
the continuum version of eq. (2) has, among other solutions,
the sine-Gordon soliton\cite{rajaraman82}, given by
\begin{equation}
\label{solitoneq}
\varphi_{s}(x, t) \sim 4 \tan^{-1}\left[ \exp \left\{ (x - v_v t ) /
\sqrt{ \eta_J^2 - \beta v_v^2 },
 \right\} \right]
\end{equation}
where $v_v$ is the velocity.
The phase in this soliton rises from $\sim 0$ to $\sim 2\pi$ in
a width $d_k \sim \sqrt{ \eta_J^2 - \beta v_v^2 }.$

Fig.\ 3 shows the local phase
difference $\varphi(x, t)$ for the Josephson ladder at two
currents in the flux-flow regime
($I/I_c = 0.22$ and $0.41$), as well as one in the ``whirling'' regime,
$I/I_c = 0.8$.($f=1/8$, $\beta=61$, $\eta_J^2 = 1.25$).
The first two show clear soliton-like behavior, namely,
an increase of $\varphi(x, t)$ in steps of
$\approx 2\pi$ over a time interval $d_k/v_v$, as predicted by the
sine-Gordon equation. The passage of the kink is accompanied
by ripples arising from phason excitations.   Typically, an
integer number of ripple periods is found between successive kink passages.
The local voltage
$\dot{\varphi}(0,t)$ (shown as broken lines)
in the flux-flow regime is due to such sine-Gordon solitons,
modified by coupling to phasons.
The sharp peaks in the local voltage
correspond to the vortex passage (we have confirmed this by snapshots of the
vortex motion in our simulations), while the exponentially decaying
smaller peaks correspond to the phasons which couple to the vortex.
In the whirling regime at $I/I_c=0.8$,
$\langle V \rangle \propto I$, and the local voltage oscillates sinusoidally in
time.
Each oscillation period
again corresponds to the passage of a vortex through a given
junction, as can be confirmed by the fact that the $v_v$ thus found is
consistent with the independently computed $\langle V \rangle$, via the
Josephson relation.

The step positions in Fig.\ 2 are determined by a locking of
the vortex motion to the phasons.  The phason dispersion relation
is determined by linearizing the left-hand side of eq.~(2)
with $i=0$.  The result is
$\varphi_m(x,t) \propto \exp(-t/2\beta)e^{i(k_m x - \omega_m t)}$,
where $\omega_m = \pm \sqrt{ 1 + 4 \eta_J^2 \sin^2(k_m/2) } / \sqrt{ \beta}$,
as obtained previously by several groups\cite{ustinov93,strogatz95}.
The allowed wave vectors $k_m$ are determined by periodic boundary
conditions: $k_m = 2\pi m / N, m=0,1,2,3,4,\ldots$.
To obtain the locking condition, note that
the vortex circulates the ladder with frequency
$\omega_v = 2\pi v_v / (Na).$   A resonance will occur if there are an
integer number of phason cycles per vortex cycle.  This condition gives
$\omega_m = n \omega_v$with $n=1,2,3,\ldots$, or
\begin{equation}
\label{conditioneq}
{1 \over \sqrt{\beta }  \left< V \right> } =\frac{ n}
{\sqrt{ 1 + 4 \eta_J^2 \sin^2 (\pi m /N)}}.
\end{equation}

All the voltage steps we have found in Fig.\ 2 satisfy this condition.
At $\eta_J = 0.71,$ for example, we identify resonances in the
range $3 < n < 15,$ $m=1$, from a high-resolution $IV$ characteristic and
its derivatives.  Presumably, the resonances corresponding to higher n
are weaker because the phasons are damped, relaxing over a time 2$\beta$.
At larger $\eta_J$, we can identify only a few values of $n$.
The values of $n$ for each step can also be found by enumerating
the number of phason wavelengths between successive vortex passages,
as in Fig.\ 3.
In the inset of Fig.\ 2 we compare the positions of the steps thus located to
the
predictions of eq. (4).  In all cases, only the
$m=1$ mode is necessary to account for the resonances.

The transition to the resistive state
(``Region II'') occurs at $n_{min}= 4,2,2,1$ for $\eta_J^2 =0.5,1.25, 2.5, 5.$
This can  also be understood from the kink-phason resonance picture.
To a phason mode, the passage of a kink of  width $d_k$ will appear
like the switching on of a step-like driving current over a time of
order $d_k/v_v$.   The kink
will couple to the phasons only if
$d_k/v_v \ge \pi/\omega_1,$ the half-period of the
phason, or equivalently
\begin{equation}
\label{resonanceeq}
{1 \over \sqrt{\beta} v_{v}} \ge { \sqrt {1 + \pi^2} \over \eta_J} =
{3.3 \over \eta_J}.
\end{equation}
This condition agrees very well with our numerical
observations, even though it was obtained by considering soliton solutions
from the continuum sine-Gordon equation.

The fact that the voltage in regime I is approximately linear in $f$
can be qualitatively understood from the following argument.
Suppose that
$\varphi$ for $Nf$ fluxons can be approximated as
a sum of well-separated solitons, each
moving with the same velocity and
described by $\varphi(x,t)=\sum_{j=1}^{Nf}\varphi_j$, where
$\varphi_j = \varphi_s(x-x_j,t)$.
Since the solitons are well separated, we can use following properties:
$\sin[ \sum_j \varphi_j ] = \sum_j \sin \varphi_j$ and $\int \dot{\varphi_j}
\dot{\varphi_i} dx \propto \delta_{ij}.$ By demanding that the energy
dissipated
by the damping of the moving soliton be balanced by that the driving current
provides($\propto \int dx i\dot{\varphi}(x)$), one can show that the $Nf$
fluxons should move with the same velocity
$v$ as that for a single fluxon driven by the same current.
In the ``whirling'' regime, the $f$-{\em independence} of the voltage can be
understood from a somewhat different argument. Here, we assume a
periodic solution of the form $\varphi = \sum_j^{Nf} \varphi_w(x - \tilde{v}t -
j/f)$
moving with an unknown velocity $\tilde{v}$ where $\varphi_w(\xi )$ describes a
whirling solution containing one fluxon.
Then using the property $\varphi (x + m/f ) = \varphi (x) + 2\pi m,$ one can
show after some algebra that
$\sin[ \sum_j^{Nf} \varphi_w(x - \tilde{v}t - j/f) ] = \sin[ Nf
\varphi_w(x-\tilde{v}t) ].$ This means that $Nf \varphi_w(x-\tilde{v}t) $ is a
solution to eq.~(2) as is $\varphi_w(x - vt).$ Finally, using the approximate
property
$\varphi_w (\xi) \sim \xi$ of the whirling state, one finds $\tilde{v} = v /
(Nf),$ leading to an f-independent voltage.

\underline{Ballistic soliton motion and soliton mass.}
A common feature of massive particles is their ``ballistic motion,''
defined as inertial propagation after the driving force has been
turned off.  Such propagation has been reported experimentally\cite{zant92}
but as yet has not been observed numerically in either square or triangular
lattices\cite{geigenmuller93,yu94}.  In the ``flux-flow'' regime at
$\eta_J = 0.71$, we also find no ballistic
propagation, presumably because of
the large pinning energies produced by the periodic lattice. (The critical
current for soliton depinning at $\eta_J = 0.71$ is about 0.2I$_c$, about twice
that calculated for a square lattice\cite{lobb}.)  However, at
$\eta_J > 1$, we do observe ballistic motion in the flux flow region I.
As an example,
Fig. \ref{ballisticfig} shows $V(t)$ and $dV/dt$ at junction 0,
for $\eta_J^2 = 5$, $I/I_c = 0.41$ (on the $n=1$ voltage step), after
the driving current is switched off at $t=0$.
The washboard-like ridges of $V$ on a decreasing background, and perhaps more
clearly, the spikes in $dV/dt$, correspond to a vortex passing through this
junction.  The increasing distance between peaks indicates that
the vortex is slowing down.  The vortex circulates at least
five times around the ring before stopping - a fact which is also verified
by direct observation of the vortex motion in real time.
Qualitatively similar behavior was also observed for slower vortices on the
lower current steps so long as $\eta_J > 1.$   This behavior can be understood
by noting that increasing $\eta_J$ {\em increases} the width of the kink $d_k$,
and thereby effectively makes the ladder seem {\em less} discrete.
The fluxon at large $\eta_J$
therefore has a much lower depinning current and can propagate
ballistically. This suggests the interesting experimental possibility that one
can tune the vortex pinning current and effective mass
by manipulation of the anisotropy in the Josephson coupling strength.

We can define the fluxon mass in our ladder by equating
the charging energy $E_c = C/2 \sum_{ij} V_{ij}^2$
to the kinetic energy of a soliton of mass $M_v$:
$E_{kin} = \frac{1}{2}M_v v_v^2$\cite{geigenmuller93}.  Since
$E_c$ can be directly calculated in our simulation,
while $v_v$ can be calculated from $\left< V\right> $, this gives an
unambiguous
way to determine $M_v$.
For $\eta_J^2 =0.5,$ we
find $E_c/C \sim 110 (\left< V \right>/I_cR)^2$,
in the flux-flow regime (region I of Fig.\ 1). This gives
$M_v^{I} \sim  3.4 C\phi_0^2 /a^2$,
more than six times the usual estimate for the vortex mass in a
2D square lattice\cite{zant93}.
Similarly, the vortex friction coefficient $\gamma$ can be
estimated by equating the rate of energy dissipation,
$E_{dis} = 1/2 \sum_{ij}V_{ij}^2/R_{ij}$, to $\frac{1}{2}\gamma v_v^2$.
This estimate yields $\gamma^{I} \sim 3.4 \phi_0^2 /(R a^2)$, once again
more than six times the value predicted for 2D arrays\cite{geigenmuller93}.
This large dissipation explains the absence of ballistic motion for this
anisotropy\cite{geigenmuller93,yu94}. At larger values $\eta_J^2 = 5$
and $2.5$,
a similar calculation gives $M_v^{I} \sim 0.28$ and $0.34 \phi_0^2 /(R a^2)$,
$\gamma^{I} \sim 0.28$ and $0.34 \phi_0^2 /(R a^2) $.  These
lower values of $\gamma^{I}$, but especially
the low pinning energies, may explain why ballistic
motion is possible at these values of $\eta_J$.

To conclude, we have shown analytically
that the dynamics of an underdamped Josephson ladder
can be approximately reduced to those of a discrete sine-Gordon
equation.  As confirmation, we showed numerically that
the ladder exhibits many phenomena previously reported in
a discrete sine-Gordon system.  These include separate flux-flow and
whirling regimes in the IV characteristics,
voltage steps in the flux flow regime which arise
from locking of vortices to phason excitations on the ladder,
and ballistic vortex propagation on the steps at sufficiently high anisotropy.
It would be of interest to construct anisotropic ladders in order to verify
some of these predictions, and to attempt to extend these arguments to
2D arrays.

We are grateful for valuable conversations with A. V. Ustinov.
This work has been
supported by NSF Grant DMR94-02131 and by the Midwest Superconductivity
Consortium through DOE Grant DE-FG02-90ER-45427.
SR was supported by Ohio State University Postdoctoral Fellowship.

\begin{figure}[ht]
\caption [flux flow] {Calculated IV curves for various values of $f$
in an $8 \times 1$ ladder ring. $f \equiv B a^2/\phi_0 = 1/8$ is the
number of flux quanta per plaquette.  We use $\beta = 33$ and $\eta_J = 0.71.$
Plotted voltages are divided by $f$.
Inset: schematic of ring topology used in simulation. A uniform current is
injected into the inner grains and drawn out from the outer grains.}
\label{fluxflowfig}
\end{figure}

\begin{figure}[ht]
\caption [dvdif1_8] {Calculated IV curves for $f = 1/8$, $\beta=61$,
and several values of the anisotropy parameter $\eta_J$. The inset shows
the voltages of steps corresponding to locking number $n$ determined from
our numerical result compared to those calculated from eq.~4
(solid lines).}
\label{dvdif1_8fig}
\end{figure}

\begin{figure}[ht]
\caption [spatiotemporal] {Local phase difference $\varphi_i(t)$(solid lines)
and voltage
$V_{i}(t)$(Broken lines) for $i = 0$, $f=1/8$, $\eta_J^2=1.25$, $\beta =61$,
and
three values of  the applied
current: $I/I_c = 0.22$ and $0.41$, corresponding to
currents on voltage plateaus with $n=3, 2$; and $I/I_c = 0.8$,
corresponding to a current in the
whirling regime. Note the different scales in the axes for the whirling
regime.}
\label{spatiotemporalfig}
\end{figure}

\begin{figure}[ht]
\caption [ballistic] {Plot of
$V_0(t)$ and $dV_0(t)/dt$ at $x=0$ after driving current is turned off at
time $t=0$.  We use $\eta_J^2 = 5$, and I/I$_c$ = 0.41,
corresponding to an $n=1$
voltage plateau.}
\label{ballisticfig}
\end{figure}

\end{document}